\documentclass[prb,twocolumn,showpacs]{revtex4-1}
\usepackage{graphicx}
\usepackage{amsfonts}
\usepackage{amsmath}
\usepackage{amssymb}
\usepackage{bm}
\usepackage{subfigure}
\usepackage{hyperref}
\begin{document}
\title{Possible particle-hole instabilities in interacting type-II Weyl semimetals}
\author{Weizhu Yi}
\author{Qiu-Shi Wang}
\author{Rui Wang}
\author{Baigeng Wang}
%\author{}
%\author{}
\affiliation{National Laboratory of Solid State Microstructures and Department of
Physics, Nanjing University, Nanjing 210093, China
\\}

\date{\today }

\begin{abstract}
Type II Weyl semimetal, a three dimensional gapless topological phase, has drawn enormous interest recently. These topological semimetals enjoy overtilted dispersion and Weyl nodes that separate the particle and hole pocket. Using perturbation renormalization group, we identify possible renormalization of the interaction vertices, which show a tendency toward instability. We further adopt a self-consistent mean-field approach to study possible instability of the type II Weyl semimetals under short-range electron-electron interaction. It is found that the instabilities are much easier to form in type II Weyl semimetals than the type I case. Eight different mean-field orders are identified, among which we further show that the polar charge density wave (CDW) phase exhibits the lowest energy. This CDW order is originated from the nesting of the Fermi surfaces and could be a possible ground state in interacting type II Weyl semimetals.
\end{abstract}

\maketitle
 \section{Introduction}
Topological semimetals have drawn enormous attention recently due to their interesting topological behavior and potential application in quantum transport\cite{C. L. Kane,H. Zhang,J. C. Y. Teo,X. L. Qi,L. Fu} Among them, Weyl semimetals (WSMs), which are featured by gapless nodal crossing points that mimic Weyl fermions in high energy physics \cite{Wan,H. Weng,Ari Turner,Ryo,W.Witczak-Krempa,GangChen}, have been extensively investigated \cite{K. Y. Yang,D. E. Kharzeev,S. Parameswarn,A. A. Zyuzin,P. Goswami}. The linear dispersion between two non-degenerate band ensures even numbers of gapless nodes in energy spectrum, which are source and drain of Berry flux in lattice momentum space. These three-dimensional (3D) semimetals are reminders of the one-dimensional Adler-Bell-Jackiw anomaly\cite{T. Meng,K. Y. Yang, H. B. Nielsen,Z. Wang,C. X. Liu,V. Aji,Pavan Hosur}, and exhibit the chiral anomaly which is further responsible for the observed negative magneto resistance in TaAs \cite{H. Weng,Chenglong Zhang,Yaojia Wang}. The standard Weyl semimetals are isotropic in their spectrum and respects the Lorentz invariance. However, in solid state physics, the Lorentz invariance can be readily broken under perturbations. Breaking the Lorentz symmetry in Weyl semimetals leads to tilted Weyl cones. When the tilting becomes severe enough, the Weyl nodes, the Fermi point for zero chemical potential, will be replaced by finite Fermi pockets. These tilted Weyl semimetals, characterized by an electron and a hole Fermi pocket, are termed as type II WSMs \cite{A. A. Soluyanov,K. Koepernik,Y. Sun,G. Auts}.

Intuitively, the overtilting of a Weyl cone only introduces anisotropic spectrum to the standard cases. From the perspective of Hamiltonian, the tilting can be described by a diagonal term $\omega\sigma^0$ in spin basis. Therefore, even though the energy eigenvalue is altered the eigenstate and hence the corresponding Berry curvature still remains the same. Followed from this simple argument, it is expected that, despite the anisotropy, the main topological behavior such as the negative magneto-resistance should not be qualitatively different from the type I WSM. However, the significant difference between type I and type II WSM is the shape of Fermi surface\cite{A. A. Soluyanov,M. O. Goerbig}. It is well known that the shape of Fermi surface and the density of states at Fermi energy are closely related to general susceptibilities under different perturbations. An natural and important perturbation on the single-particle picture of WSMs is the electron-electron interaction, hence it is interesting to consider the question, i.e., what is the possible instability of the type II WSM under interaction, and how is it different from the type I WSM.

The instability of type I WSM has been investigated by previous works\cite{Fabrizio Detassis}. Firstly, it was found that the short-range interaction is irrelevant due to the linear dispersion and the vanishing density of states at the Weyl point \cite{Joseph Maciejko}. This means that weak and well-screened interaction can be neglected, accounting for the stability of type I WSM in experiment. Secondly, the strong interaction effect is studied. Due to the vanishing density of states and the linear dispersion, a critical interaction threshold is required to induce different phases. Various insulating phases where the Weyl nodes are gapped out are identified by H. Wei et al., among which the chiral excitonic insulator is found to be a promising candidate \cite{Wei2012}. In the type II WSM, the key difference is the emergence of electron and hole pockets and a finite density of states at Fermi energy. Even though some attempts have been made in 2D cases \cite{Shun-Qing Shen}, a systematic investigation of the strong interacting 3D type II WSM is still lacking.

In this work, we first use a perturbative renormalization group method to identify possible flowing tendency of interaction constant under different energy scale. It is found that a relevant contribution to the renormalization of interaction will arise in the second order calculation, which indicates a tendency for instability. Then, we go beyond perturbation theory, and apply a mean-field theory to study the type II WSM with interaction. By symmetry analysis, we find that eight possible states are possible which all spontaneously break the rotation symmetry of the parent state. Through self-consistent calculation, two interesting properties that are unique in type II case are extracted. First, the instabilities with nonzero mean-field order parameter are found to exist at least down to $g=0.1$ (where $g=0.47$ for momentum cutoff $\Lambda=1$ in the type I WSM). This means that the symmetry breaking orders are much more easily to be formed due to the finite density of states and emergence of Fermi pockets in type II WSM. Second, the polar charge density wave (CDW) phase is mostly energetically favored among all eight possible states, as a result of Fermi surface nesting between those around the two tilted Weyl cones.

The remaining sections of this work is organized as follows. In Sec.II, we perform a perturbative renormalization group calculation of the flowing equations of different parameters. In Sec. III, a self-consistent mean-field theory is constructed. Using this formalism, eight different possible candidate phases are introduced and compared in terms of their energetics. Finally, concluding remarks are included in Sec.IV.

\section{Perturbative Renormalization Group Analysis}
Before consider a type II WSM, we first study a single Weyl cone. The minimal low-energy effective model describing an interacting tilted Weyl cone reads as,
\begin{equation}
	H=\int d^{3}k\psi ^\dagger \hbar v_{F}(\mathbf{k}\cdot \mathbf{\sigma }+\boldsymbol{\omega}\cdot \mathbf{k})\psi +g\int d^{3}k\psi^\dagger\psi ^\dagger\psi \psi,
	\label{equ1}
\end{equation}	
where $\mathbf{\omega }$ is a dimensionless vector along which the Weyl cone is tilted. For $|\omega|>1$, electron and hole pockets take place, as is the case for the type-II WSM. We use a Hubbard type interaction in the second term to describe the short-range local Coulomb interaction in the WSM, where $g$ is a positive coupling constant.
\begin{figure}[tbp]
\includegraphics[width=2.4in]{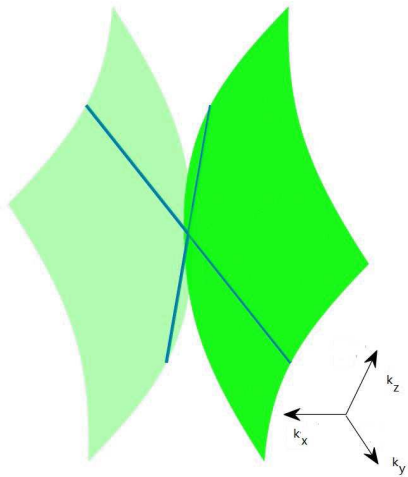}
\caption{The Green surfaces are the Fermi surfaces. The x-y plane intersects with the Fermi surface at the two blue lines.}
\end{figure}

\begin{figure}[htbp]
		\centering
		\subfigure[]{
			\label{Fig.sub.1}
			\includegraphics[width=0.36\textwidth]{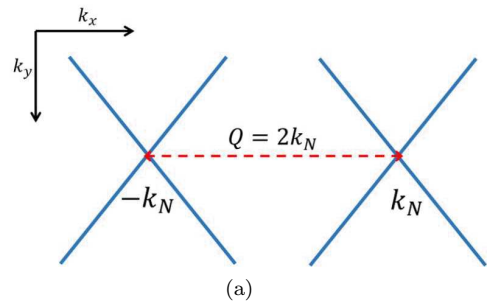}}
		\subfigure[]{
			\label{Fig.sub.2}
			\includegraphics[width=0.40\textwidth]{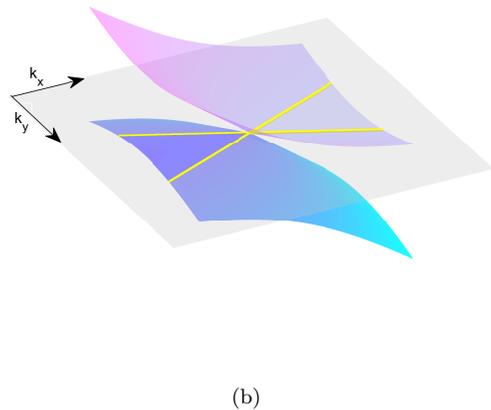}}
		\caption{(a)The schematic plot of the Fermi surfaces($k_{z}=0$) around two Weyl points, (b)The dispersion of Type-II Weyl semimetals with $k_{z}=0$ and intersects with the zero-energy plane at the yellow crossed lines which is also the Fermi surfaces($k_{z}=0$) around a single Weyl point.  }
		\label{Fig.lable}
	\end{figure}

To study the effect of the interaction, we first employ a perturbative renormalization group (RG) to study the flowing behavior of coupling constants. Since Eq.\eqref{equ1} is a low-energy effective model, a momentum cutoff $\Lambda$ is implicit in the sum of $\mathbf{k}$. In the standard RG treatment, integration of momentum space of fast mode is performed iteratively, gradually reducing $\Lambda$ down toward the Fermi surface. As is shown in Fig.1, for fixed $k_z$, the Fermi surface around each Weyl cone are crossing lines that separates the electron and hole pocket. For each RG step, one integrates out the shell region in Fig.2, arriving at a  coarse-grained model with renormalized coupling constants. It is also worthwhile to note that, when a realistic type II WSM state is considered, the Fermi surfaces around each Weyl node are nested through umklapp momentum $Q$. This would indicate possible instability via internode coupling. In this section, we neglect such effect and leave it to the content below.

As a perturbative theory, we calculate the Feynman diagrams that contribute to the renormalization of our parameters, i.e., $v_F$, $\omega$, and $g$.
First, the propagator of free particle in type-II Weyl Semimetals reads as,
\begin{equation}
\begin{split}
	G_{0}\left ( i\omega ,\mathbf{k} \right )&=\frac{1}{i\omega -v_{x}k_{x}\sigma ^{1}-v_{y}k_{y}\sigma ^{2}-v_{z}k_{z}\sigma ^{3}-\boldsymbol{\omega}\cdot \mathbf{k}\sigma ^{0}}\\
&=\frac{1}{i\omega -v_{F}\mathbf{k}\cdot \mathbf{\sigma}-\omega v_{F}k_{x} },
\end{split}
\end{equation}
where we have chosen the vector to be $\boldsymbol{\omega}=\omega\mathbf{e}_x$, without losing any generality. The full propagator of electrons is given as
\begin{equation}
	G^{-1}=\left ( G^{0} \right )^{-1}-\Sigma,
\end{equation}
where $\Sigma$ is the self-energy received from the interaction correction. In the RG analysis, the self-energy $\Sigma$ is calculated within the fast-mode momentum shell, which will in turn renormalize the parameters $v_F$ and $\omega$.  The interaction vertex is represented by the Feynmann diagram in Fig.3(b). Then, at tree level, the self-energy correction is calculated by  Fig.3(a), leading to
	\begin{figure}[htbp]
		\centering
		\subfigure[]{
			\label{Fig.sub.1}
			\includegraphics[width=0.2\textwidth]{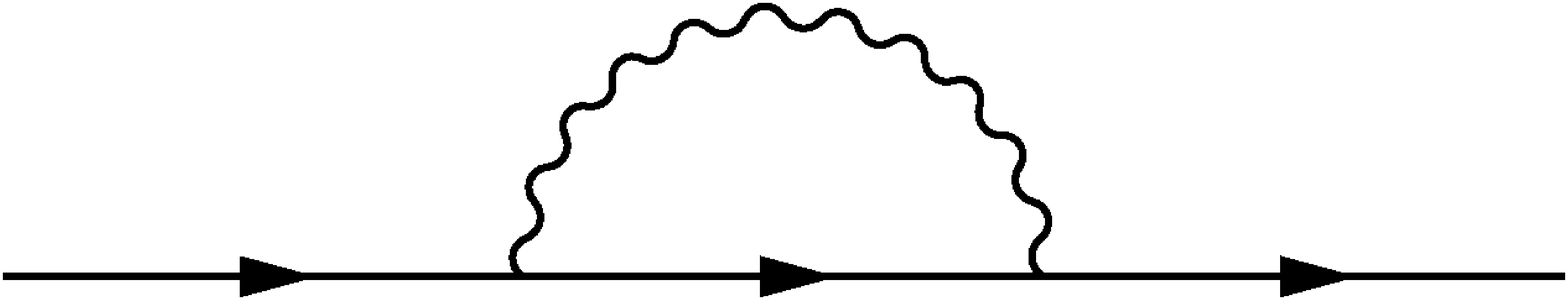}}
		\subfigure[]{
			\label{Fig.sub.2}
			\includegraphics[width=0.2\textwidth]{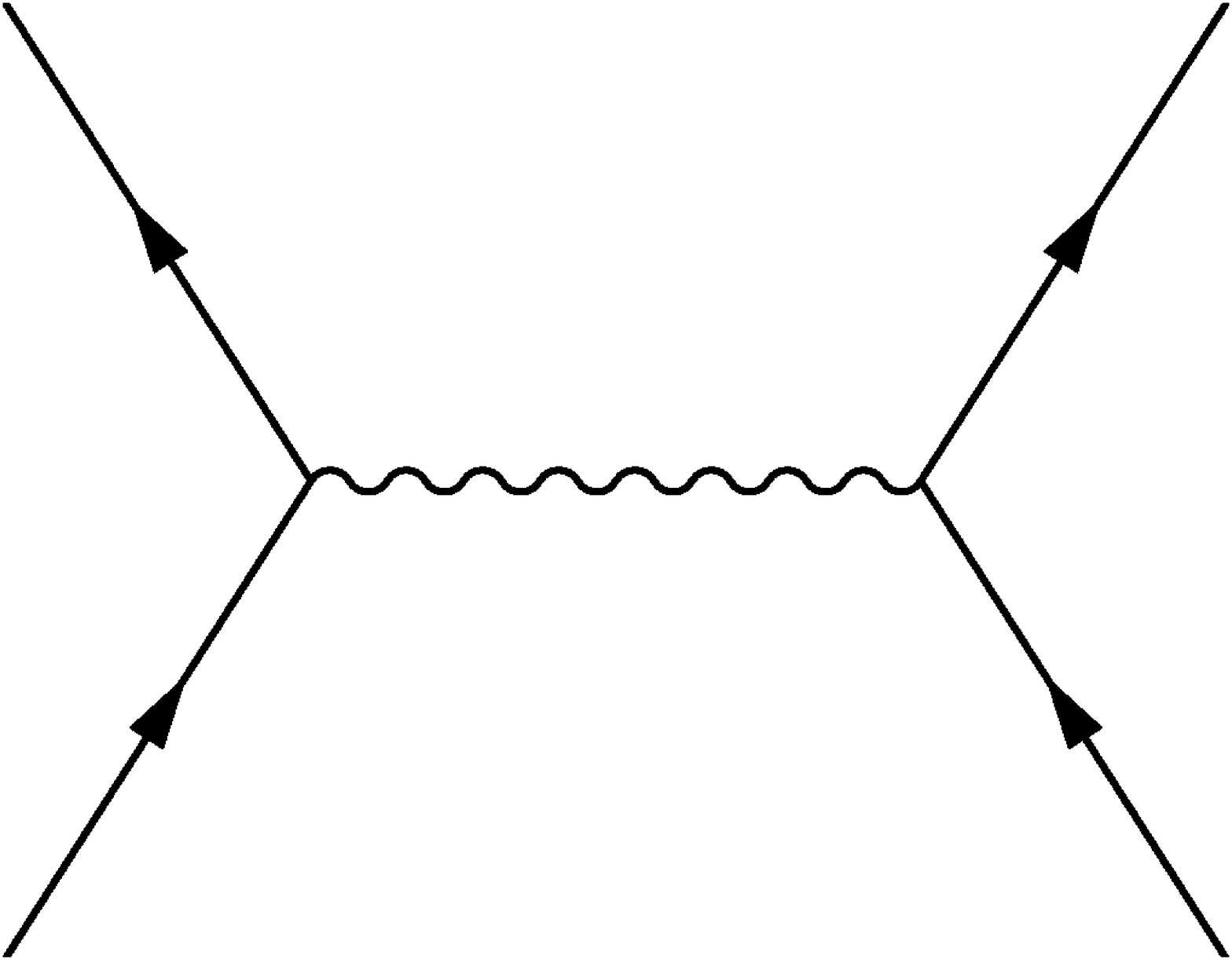}}
		\caption{Tree level correction to interaction}
		\label{Fig.lable}
	\end{figure}

	\begin{equation}
	\Sigma ^{\left ( 1 \right )}\left ( \mathbf{k},i\omega  \right )=\frac{1}{\beta }\sum_{\omega '}\int \frac{d^{3}p}{\left ( 2\pi\right )^{3}}G_{0}\left ( \mathbf{p},i\omega +i\omega ' \right )V\left ( \mathbf{k-p} \right)
	\end{equation}
After substituting $G_{0}$ and $V$, and transforming the summation into integration over $\mathbf{p}$, we arrive at
	\begin{equation}
	\Sigma ^{\left ( 1 \right )}\left ( \mathbf{k},\omega  \right )=-\frac{g}{2v_{F}}\int \frac{d^{3}k}{(2\pi )^{3}}\frac{\mathbf{k\cdot \sigma }}{\left | \mathbf{k} \right |}=0.
	\end{equation}
The zero self-energy indicates that the tilting parameter $\omega$ and the Fermi velocity $v_F$ does not flow with energy scale for the instantaneous local Coulomb interaction considered here.

Now let us consider the renormalization of the interaction vertex up to one-loop order. Three
Feynman diagrams are encountered, which are shown in  Fig.4(a)-(c).
	\begin{figure}[htbp]
		\centering
		\subfigure[]{
			\label{Fig.sub.1}
			\includegraphics[width=0.2\textwidth]{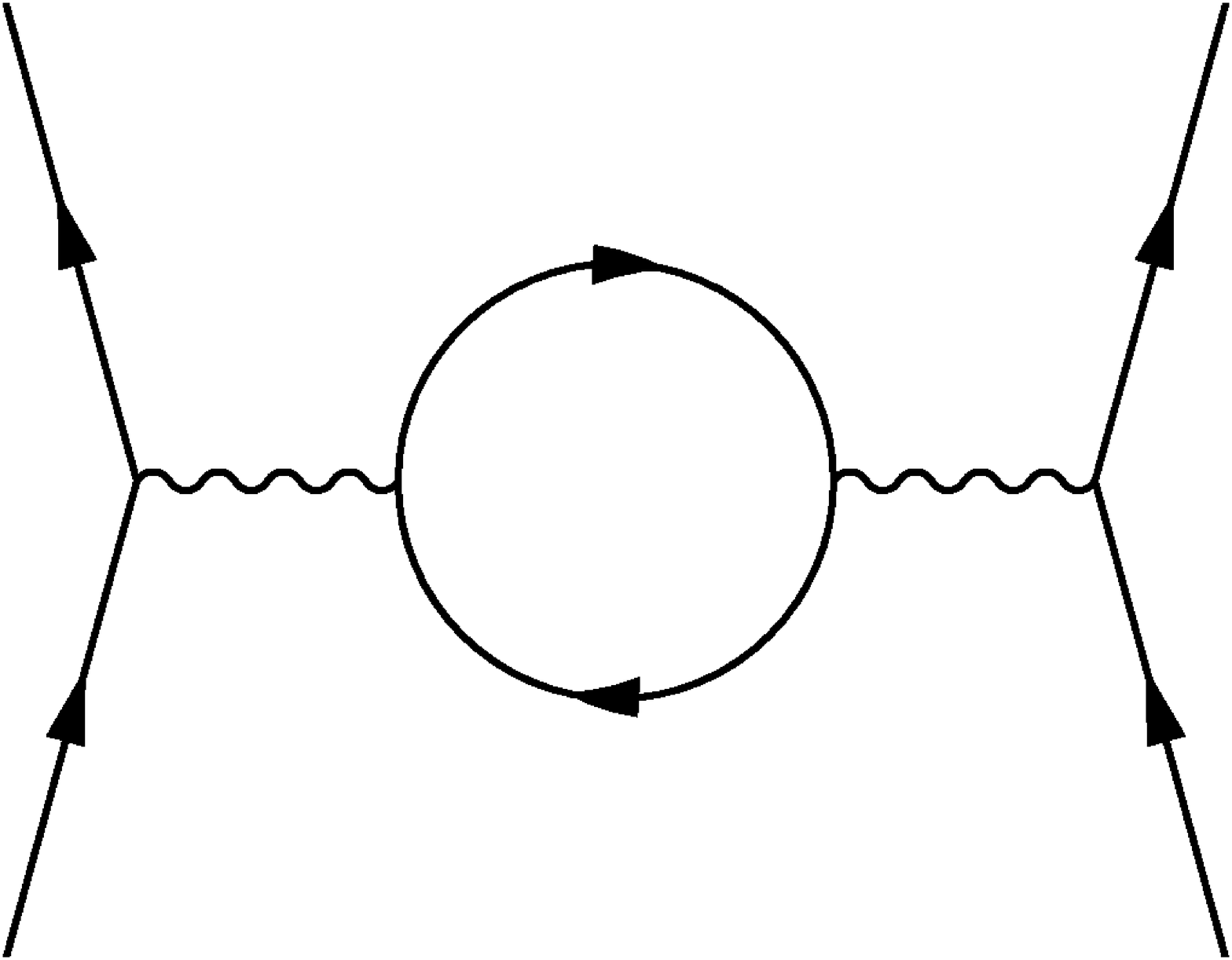}}
		\subfigure[]{
			\label{Fig.sub.2}
			\includegraphics[width=0.2\textwidth]{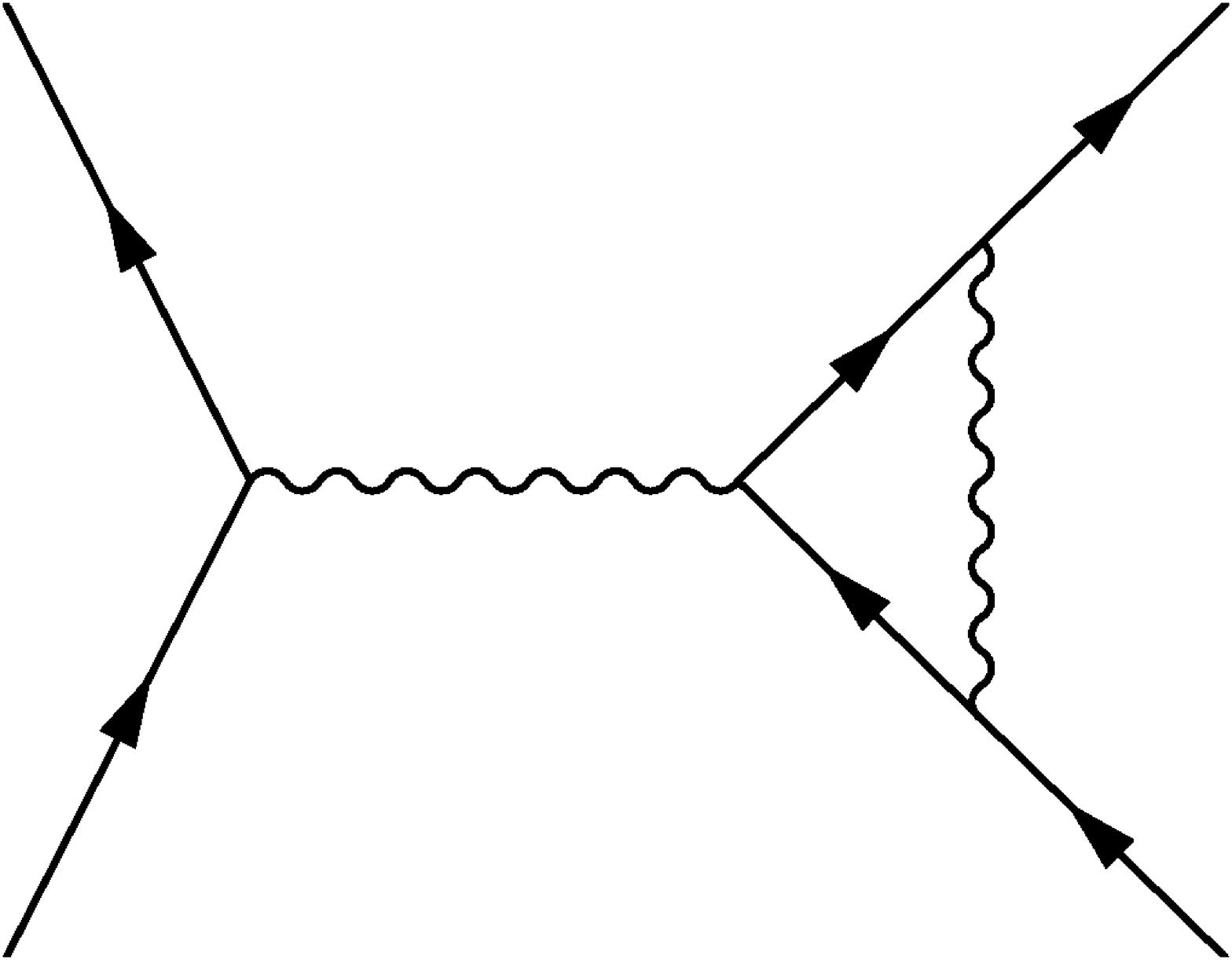}}
		\subfigure[]{
			\label{Fig.sub.3}
			\includegraphics[width=0.2\textwidth]{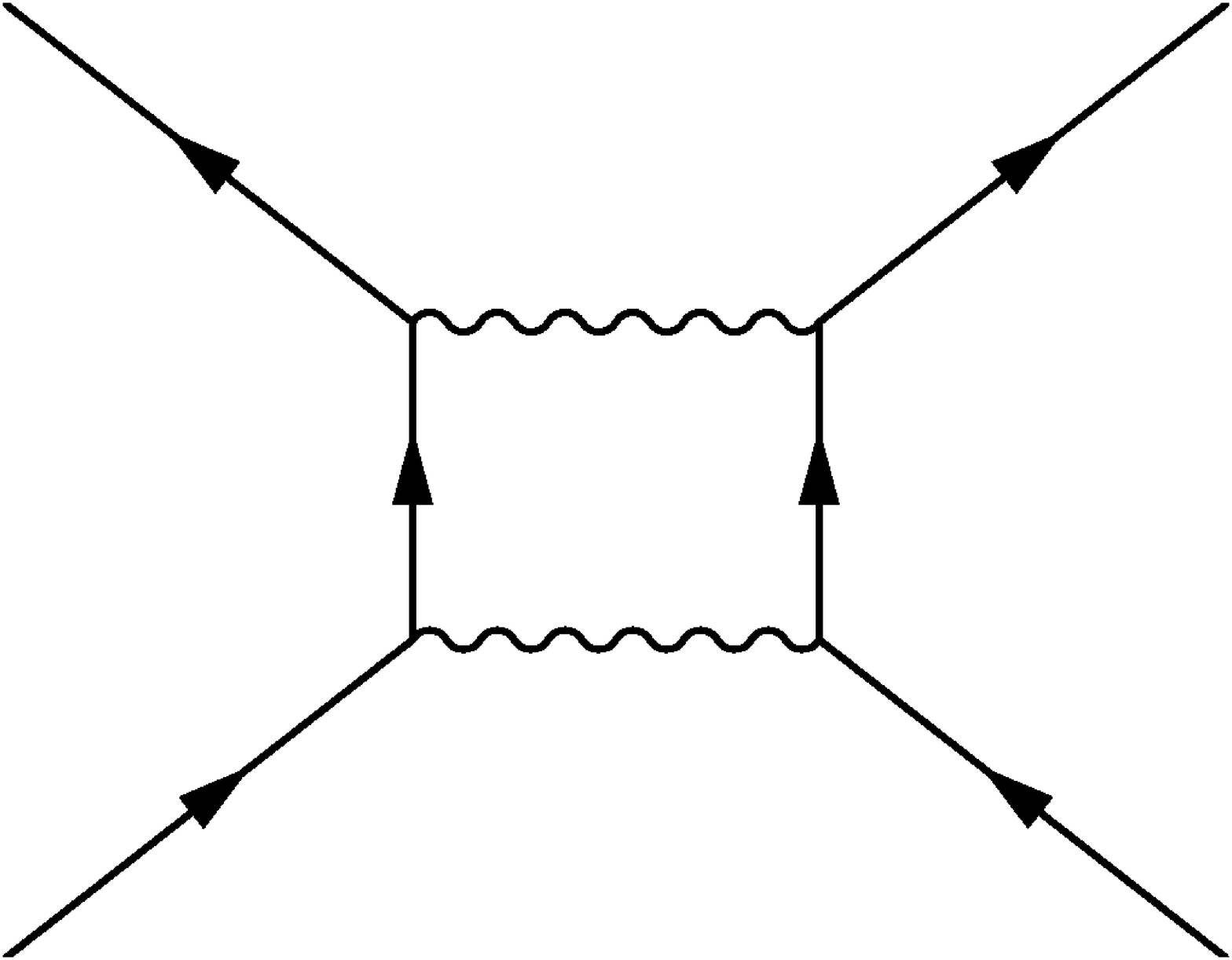}}
		\caption{One-loop correction to interaction}
		\label{Fig.lable}
	\end{figure}
Any indefinite momentum and frequency in the closed loop should be integrated, leading to the following three integrals $I_{A,B,C}^{(1)}$, which respectively reads as
\begin{equation}
\begin{split}
	I_{A}^{(1)}\left ( i\nu ,\mathbf{q} \right )
&=-g^{2}\int \frac{d\omega }{2\pi }\int \frac{d^3p}{ ( 2\pi )^{3}}\\
&\times\textup{Tr}\left [ G\left ( \mathbf{p+q},i\omega +i\nu  \right ) G\left ( \mathbf{p},i\omega  \right )\right ],
\end{split}
\end{equation}
\begin{equation}
	I_{B}^{(1)}\left ( i\nu ,\mathbf{q} \right )=g^{2}\int \frac{d\omega }{2\pi }\int \frac{d^3p}{ ( 2\pi)^{3}}\left [  G\left ( \mathbf{p},i\omega  \right )G\left ( \mathbf{p+q},i\omega +i\nu  \right )\right ],
	\end{equation}
and
\begin{equation}
\begin{split}
	I_{C}^{(1)}\left ( i\nu ,\mathbf{q} \right )&=g^{2}\int \frac{d\omega }{2\pi }\int \frac{d^3p}{ ( 2\pi)^{3}}\\
&\left [  G\left ( \mathbf{k+q-p},i\eta +i\nu-i \omega  \right )G\left ( \mathbf{p+k'},i\omega +i\eta ' \right )\right ],
\end{split}
	\end{equation}
%
%	then
%	\begin{gather}
%	I_{A}^{(1)}\left ( i\nu ,\mathbf{q} \right )=-g^{2}\int \frac{d\omega }{\pi }\int \frac{d^3p}{ ( 2\pi )^{3}}\frac{-\omega (\omega +\nu )+v_{F}^{2}\mathbf{p\cdot (p+q)}}{\left [ (\omega +\nu )^{2}+v_{F}^{2}\mathbf{ \left | p+q \right |^{2}} \right ]\left [ \omega ^2 + v_{F}^{2}\mathbf{\left | q \right | }^{2}\right]}  \notag \\
%	=g^{2}\int_{0}^{1}dx\int \frac{d^3p}{ ( 2\pi )^{3}}\frac{x(1-x)v_{F}^{2}\mathbf{q}^{2}}{\left [v_{F}^{2}\mathbf{p}^{2}+x(1-x)(v_{F}^{2}\mathbf{q}^{2}+\nu ^{2}) \right ]^{3/2}}
%	\end{gather}
where all the integrals of momentum are performed within the fast-mode shell. Summing over the three diagrams, one can finally find out that the coupling constant $g$ is renormalized to
\begin{equation}\label{eqq1}
g^{\prime}=g+\lambda(\mathbf{k},\mathbf{k}^{\prime},\mathbf{q}) g^2 dl,
\end{equation}
where $l$ is the RG flowing parameter, and
\begin{equation}\label{eqq2}
 \lambda(\mathbf{k},\mathbf{k}^{\prime},\mathbf{q})=\frac{(\mathbf{q+k+k'})^2}{v_F} \frac{\pi w^2 (w+2) (2 w+3)}{ 6(w-1) (w+1)^3}.
\end{equation}
The above equation only applies for the type II case with $|\omega|>1$, due to the modification of the integral condition in this case. It is obvious from Eq.\eqref{eqq2} that for $|\omega|>1$, $\lambda$ is a positive parameter, therefore indicating a relevant contribution in the one-loop order. From the above renormalization of the coupling constants, two conclusions can be drawn. First, the simple Hubbard type local interaction acquires momentum-dependence.  Second, due to $\lambda>0$ for the type II case, the second order correction in Eq.\eqref{eqq1} suggests a tendency for instability.

 \section{Mean-field analysis}
Based on the RG analysis above, it is known that up to second order of $g$,  a relevant term will emerge in the RG flow of the interaction $g$, suggesting a tendency for instability.  In this section, we further consider the effect of repulsive interactions and study the what is the specific instability that is energetically favorable in type II WSMs. In order to do so, we focus on the minimum model with two symmetrically tilted Weyl cones, namely, the Weyl cones enjoy opposite tilting direction with opposite chirality. The schematic plot of the energy spectrum is shown in Fig.5. \par

%\begin{figure}[htbp]
%	\centering\includegraphics[width=3.5in]{3.png}
%	\caption{Type-II Weyl Semital}\label{Fig.lable}
%\end{figure}

To describe such two Weyl nodes located at $ \mathbf{ k_N }$ (labeled by R) and $ -\mathbf{ k_N }$ (labeled by L) with chirality +1 and -1, one can write down the Hamiltonian as
\begin{equation}
H_0=\pm \hbar v_F\sum_{\mathbf{k}}\psi ^{\dagger }_{\mathbf{k}\alpha }(\mathbf{\sigma _{\alpha \beta }+
	\omega \cdot \mathbb{I} _{\alpha \beta }})\cdot (\mathbf{k\mp k_N})\psi_{\mathbf{k}\beta }
\end{equation}
where $ v_F $ is the Fermi velocity and $ \mathbf{\sigma} $ is the Pauli matrices denoting the spin degrees of freedom. The conduction (valence) band at the R node has its spin parallel (anti-parallel) to $\mathbf{k\mp k_N}  $, $\boldsymbol{\omega}$ is the tilted vector
 and $|\omega|>1$ for the type II WSM.

For the interaction terms, we take the form
\begin{equation}
V=\sum_{\sigma ,\sigma '}\sum _{\mathbf{k,k',q}}V\left ( \mathbf{q} \right )\psi ^{\dagger} _{\mathbf{k'+q,\sigma}}\psi ^{\dagger} _{\mathbf{k'-q,\sigma^{\prime} }}\psi _{\mathbf{k',\sigma '}}\psi  _{\mathbf{k,\sigma }}
\end{equation}
where $\mathbf{k,k'}$ are are the scattering momenta of the incoming electrons (holes) and $ \mathbf{k'+q,k-q} $ are the scattering momenta of the outgoing electrons (holes) respectively, with a exchange moment of $\mathbf{q}$ through interaction.. Since there are two Weyl nodes and the electron operators in the interaction can come from the excitation around both nodes, there are 16 different scattering terms in total. After restricting to the low energy sector, the conservation of energy and momentum will rule out most of the terms, leading to 6 possible combinations of scattering processes between two Weyl nodes, i.e., (L,L,L,L),(R,R,R,R), (L,R,L,R), (L,R,R,L), (R,L,L,R), and (R,L,R,L). The other terms, for example,(L,L,R,R), can be neglected due to the violation of the conservation of momentum.

In order to study possible instabilities in the particle-hole channel, it is convenient to firstly express the interaction $V$ in the diagonal basis of unperturbed Hamiltonian $H_0$. Hence, we first solve the Shrodinger equations,
\begin{equation}\label{eqa1}
H_{0,\pm }^{L,R}\chi_{\mathbf{q},\pm }^{L,R}=\pm E_0^{L,R}\chi_{\mathbf{q},\pm }^{L,R}
\end{equation}
to obtain the corresponding eigenvectors. Then we define $\psi  _{\mathbf{q,\sigma },\pm }^{L,R}=\chi_{\mathbf{q},\pm }^{L,R}c_{\mathbf{q},\pm }^{L,R}$ as the fermionic fields of the conduction bands and valence bands, where $\chi$ is the spinor denoting the eigenvectors, which is obtained as,
\begin{equation}
\begin{split}
&\chi_{\mathbf{q},+(-)}^{R(L)}=\binom{\cos\frac{\theta }{2}e^{-i\phi }}{\sin\frac{\theta }{2}}\\
&\chi_{\mathbf{q},-(+)}^{R(L)}=\binom{\sin\frac{\theta }{2}e^{-i\phi }}{-\cos\frac{\theta }{2}}.
\end{split}
\end{equation}
In Eq.\eqref{eqa1} , $c_{\pm}$ is the annihilation operator of the conduction and valence band electrons, respectively. The angles $ \theta $ and $ \phi $ are the polar and azimuthal angles of the momentum vector $\mathbf{q}$.
In order to simplify the expression of the interaction, we introduce a new set of orthogonal coordinate system in the momentum space as that in Ref.\cite{Wei2016}, namely, two additional vectors $\mathbf{\hat{e}_1},\mathbf{\hat{e}_2}$ are introduced
that are perpendicular to the momentum $\mathbf{k}$, forming a three-dimensional right-handed coordinate.

Let us then consider the scattering process in the low-energy window. In order to satisfy the energy conservation in addition to the momentum conservation, all possible momentum exchange in the scattering process can be found as $\mathbf{k-k'}  $,$\mathbf{k-k'-2k_N}  $ or $\mathbf{2k_N}$. This restriction leads to several allowed scattering channels, and the analysis here is similar to that in the RG approach where only three scattering channels are marginal in the tree level. In this way, the interaction terms are further simplified in the low-energy window. After a straightforward calculation, the reduced interaction is obtained as,

\begin{equation}\label{eqa2}
  V=V_{inter}+V_{intra},
\end{equation}
where
\begin{equation}\label{eqa3}
\begin{split}
  V_{inter}&=\sum _{\mathbf{k,k',\pm }}
\left [2V\left ( 2\mathbf{k_N}\right )-V\left ( \mathbf{k-k'} \right ) \left ( \mathbf{\hat{k}\cdot \hat{k}'}+1 \right )\right ]\\
&\times c^{L \dagger }_{\mathbf{k},\pm }c^{R\dagger }_{\mathbf{k'},\mp }c^{ L}_{\mathbf{k'},\pm }c^{R }_{\mathbf{k},\mp },
\end{split}
\end{equation}
and
\begin{equation}\label{eqa4}
\begin{split}
  V_{intra}&=-\sum _{\mathbf{k,k',\pm }}\frac{\mathbf{\hat{u}\cdot \hat{u}'^{\ast }+\hat{u}'\cdot \hat{u}^{\ast }}}{4}
V\left ( \mathbf{k-k'} \right ) \\
&\times\sum _{R,L}c^{\tau \dagger }_{\mathbf{k},\pm }
c^{\tau \dagger }_{\mathbf{k'},\mp }c^{\tau }_{\mathbf{k'},\pm}c^{\tau}_{\mathbf{k},\mp }\\
&-\sum _{\mathbf{k,k',\pm }}\frac{\mathbf{\hat{u}\cdot \hat{u}'+\hat{u}^{\ast }\cdot \hat{u}^{'\ast }}}{2}
V\left ( \mathbf{k-k'-2k_N} \right )\\
&\times c^{L \dagger }_{\mathbf{k},\pm }c^{R\dagger }_{\mathbf{k'},\mp }c^{ R}_{\mathbf{k'},\pm }
c^{L  }_{\mathbf{k},\mp},
\end{split}
\end{equation}
where $\mathbf{\hat{u}}=\mathbf{\hat{e}_1}+i\mathbf{\hat{e}_2}$. The above interaction has been derived by Ref.\cite{Wei2012}. Here, we are going to show that, the interaction, when present in type II WSM, will exhibit different effects on the semimetal phase, as a result of the finite Fermi surface.

The above interaction is dependent on three vectors $ \mathbf{\hat{k}},\mathbf{\hat{e}_1},\mathbf{\hat{e}_2} $. Each vector couples to an operator in the particle hole channel. It is then expected that $V_{intra}$ favors the instability towards excitonic insulator with intra-nodal order parameter and  $V_{inter}$ promotes the instability towards charge density wave with the inter-nodal order parameter. As will be discussed in detail below, due to the vectors emerged in the interaction, the ground state would spontaneously break the rotation invariance, leading to many different orders in the mean-field level. As has been discussed in the case of $^3\mathrm{He}$ superfluid \cite{Volovik}, the breaking of symmetry always leads to two types of configuration of order parameter which are the energy minimum, i.e., the chiral and polar phases. According to our analysis below, we are going to show that there are totally 8 possible states that are promising candidate for the ground state, namely, two CDW orders (1a,1b), four type I EI orders, EI-1(2a,2b,2c,2d), and two type II EI orders, EI-2(3a,3b).

Let us firstly focus on the inter-nodal instabilities that establish ordering at a nesting vector $ 2k_N $. The important nature in $V_{inter}$ is the p-wave like feature of the interaction. To study the vectorial leading term, we take the form of effective potential as
\begin{equation}
V_1=-\frac{g}{\Omega }\sum _{\mathbf{k,k',\pm }}
\left [ ( \mathbf{\hat{k}}c^{L \dagger }_{\mathbf{k},\pm }c^{R }_{\mathbf{k},\mp })\cdot(\mathbf{\hat{k'}}c^{R\dagger }_{\mathbf{k'},\mp }c^{ L}_{\mathbf{k'},\pm })\right ].
\end{equation}
The above interaction is formally similar to that of the interaction in $ ^3 $ He. However, in the later case, it is the condensation in the particle-particle channel that leads to chiral superfluidity. Here, for the WSM, it is the particle-hole channel where the instability would take place. Due to the finite momentum transfer at the nesting vector $k_N$, the order parameter developed enjoys the periodical real space dependence, and is therefore the CDW ordering.

In the mean-field treatment, we introduce the order parameter as
\begin{equation}
\mathbf{\Delta}=\frac{g}{\Omega }\left \langle \sum _{\mathbf{k'}}\hat{\mathbf{k'}} c^{\tau  \dagger }_{\mathbf{k'},\pm }c^{\bar{\tau} }_{\mathbf{k'},\mp }\right \rangle.
\end{equation}
Similar to the $ ^3 $ He superfluid, the symmetry analysis (cite) indicates two possible states, i.e., Chiral CDW (1a) and Polar CDW (1b). They are p-wave CDW with the following order parameter
\begin{gather}
\boldsymbol{\Delta}_{1a}=\Delta_c(\frac{\hat{x}+i\hat{y}}{\sqrt{2} }),\notag \\
\boldsymbol{\Delta}_{1b}=\Delta_p\hat{z}.
\end{gather}
Following the standard mean-field procedure, the self-consistent equation is
obtained as
\begin{equation}
1=\frac{g}{\Omega }\sum _{\mathbf{k}}\frac{\left | \hat{ \boldsymbol{\Delta}}\cdot \hat{\mathbf{k}}\right |^{2}}{2E_{\mathbf{k}}}\textup{tanh}\frac{\beta E_{\mathbf{k} }}{2}
\end{equation}
where
\begin{equation}
E_{\mathbf{k}}=\sqrt{\left ( \hbar v_{F} \right )^{2}\left ( |\mathbf{k}|+\omega k_{z} \right )^2+\left |\boldsymbol{\Delta \left ( \mathbf{k} \right )\cdot \hat{k}} \right |^{2}}
\end{equation}
where $ \mathbf{k} $ is moment relative to Weyl nodes, $ \beta=1/k_BT $ and $\hat{ \boldsymbol{\Delta}}$ is defined by $\hat{ \boldsymbol{\Delta}}= \boldsymbol{\Delta}/|\Delta|$. Since we are studying a low-energy continuum model, the sum of $ \mathbf{k} $ is restricted within an energy cutoff $ \Lambda $ around each node. After transforming the sum into the integral, one can arrive at,
\begin{equation}\label{eqa5}
\begin{split}
1&=\frac{g}{2\Omega }\int_0^{\frac{\Lambda / \hbar v_F }{w^2-1}}\int^{t^\ast }\int_0^{2\pi}d\rho dtd\vartheta\\
&\times\frac{\rho ^2w(w^2-1)\textup{cosh}t\textup{sinh}t}
{\sqrt{\rho ^2+ | \boldsymbol{\Delta \cdot \hat{k}}  |^2}}
\left |\boldsymbol{\hat{\Delta}\cdot \hat{k}} \right |^2,
\end{split}
\end{equation}
where we have made a hyperbolic transformation. In this equation, the integral limit $t^\ast$ is determined by the boundaries of two branches of hyperbolic curve determined by the energy cutoff $ \Lambda $ and  a natural momentum cutoff along z-axis $ \tilde \Lambda_0/\hbar v_F $ (see Appendix).
%Explicitly,we have
%let
%\begin{gather}
%\mu =\tilde\Lambda_0/\hbar v_F \rho -w \notag \\
%\lambda =\tilde\Lambda_0/\hbar v_F \rho +w
%\end{gather}
%if $ \mu>1 $, the integral range $  t^\ast$ have two branches contribution, for one branch, it is
%$ \left [\ln\left [ \mu -\sqrt{\mu ^2-1} \right ],
%\ln\left [ \mu +\sqrt{\mu ^2-1} \right ]  \right ] $ and the other branch $ \left [\ln\left [ \lambda -\sqrt{\lambda ^2-1} \right ],
%\ln\left [ \lambda +\sqrt{\lambda ^2-1} \right ]  \right ] $\par
%if $ -1<\mu<1 $, there is only one branch $ \left [\ln\left [ \lambda -\sqrt{\lambda ^2-1} \right ],
%\ln\left [ \lambda +\sqrt{\lambda ^2-1} \right ]  \right ] $ left\par
%if $ \mu<-1 $, there is also one branch left while it is separated into two parts, i.e.
%$ \left [\ln\left [ -\mu -\sqrt{\mu ^2-1} \right ]\right ],
%\left [\ln\left [ \lambda -\sqrt{\lambda ^2-1} \right ]\right ] $ and
%$ \left [\ln\left [ -\mu +\sqrt{\mu ^2-1} \right ]\right ],
%\left [\ln\left [ \lambda +\sqrt{\lambda ^2-1} \right ]\right ] $\par
In our numerical solution, we set $ \Lambda $ and $\Omega $ to 1 without losing any generality. After normalizing the quantities and making them dimensionless, we solve the self-consistent equation and plot the results for 1a and 1b phase in Fig.5.\par
\begin{figure}[tbp]
\includegraphics[width=3.9in]{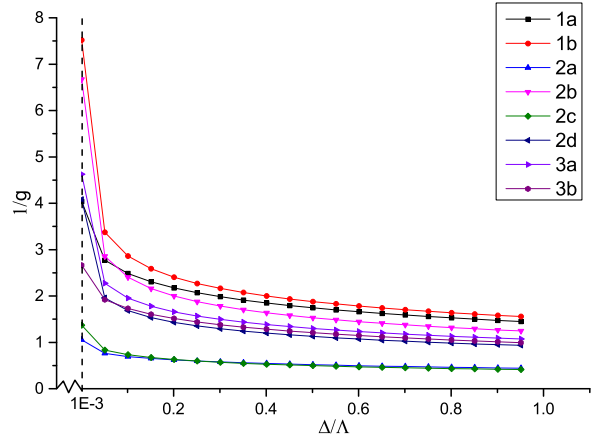}
\caption{The order parameter self-consistently calculated from the mean-field equation, versus the interaction $g$. }
\end{figure}
Several conclusions can be drawn. First, $(g,\Delta)=(0,0)$ is a solution of the above equation. This results can be understood from the analytic behavior of the above self-consistent equation. If one sets $ \Delta=0 $ in the right hand side of Eq.\eqref{eqa5}, then one can find that the integration is divergent, which then requires $g=0$. This trivial solution means that no order would be formed for zero interaction $g$. Second, the numerical calculation down to $\Delta\sim0.001$ shows that, the self-consistent equation has a nontrivial solution where $ g \sim 0.1$. Due to the limitation of precision, we did not plot the results for lower $\Delta$. However, the results already show qualitative difference between the type I WSM, where a threshold $g_c=0.47$ (for $\Lambda=1$) is required for the instabilities. Here, we find that the type II WSM is much more easily to exhibit spontaneous symmetry breaking in the particle-hole channel. This is in consistent with the fact that unlike the type I case, the type II WSM enjoys a finite Fermi surface and a nonzero density of states at the Fermi energy.

Having investigated the inter-node instability, let us now shift our attention to that due to the intra-node scattering. In the intra-node interaction in Eq.\eqref{eqa4}, one can expand the $\hat{\mathbf{u}}$ vectors and separate the interaction into two independent sectors, one only contains $ \hat{\mathbf{e}}_{1} $  while the other contains $ \hat{\mathbf{e}}_{2} $.  We first consider the $ \hat{\mathbf{e}}_{1} $  sector, which reads as,
\begin{equation}
\begin{split}
V_2&=-\frac{g}{2\Omega }\sum _{\mathbf{k,k,\pm  }}\  \left [\hat{\mathbf{e}}_{1} (c^{L\dagger }_{\mathbf{k},\mp }c^{L }_{\mathbf{k},\pm  }+c^{R\dagger }_{\mathbf{k},\mp } c^{R }_{\mathbf{k},\pm  })\right ]^{\star} \\
 &\cdot\left[\hat{\mathbf{e}}_{1}' (c^{L\dagger }_{\mathbf{k'},\mp }c^{L }_{\mathbf{k'},\pm  }+c^{R\dagger }_{\mathbf{k'},\mp } c^{R }_{\mathbf{k'},\pm  })\right ].
\end{split}
\end{equation}
Similar to the treatment in the inter-node scattering, we introduce mean-field order parameter as,
\begin{equation}
\boldsymbol{\Delta}_L+\boldsymbol{\Delta}_R=\frac{g}{2\Omega }
\left \langle  \sum _{\mathbf{k'}}\hat{\mathbf{e}}_{1}' (c^{L\dagger }_{\mathbf{k'},\mp }c^{L }_{\mathbf{k'},\pm  }+
c^{R\dagger }_{\mathbf{k'},\mp } c^{R }_{\mathbf{k'},\pm  }) \right \rangle.
\end{equation}
Before proceeding, we recall that while the vector $ \mathbf{\hat{k}}$ spans the whole unit sphere in momentum space the vector $\mathbf{\hat{e}_1}$ only spans in the southern hemisphere. This fact should not be neglected since it breaks the homogeneity of the momentum space by making the $ {\hat{z}}$ axis physcially different from the other two axes. Therefore, to completely discuss all the possible symmetry breaking phases, one has to consider two distinct cases in the polar phases, i.e., (A) a vectorial order parameter which points towards $ {\hat{z}}$ direction and (B) the order parameter whose vectorial direction is perpendicular to the $ {\hat{z}}$ direction.  For the chiral phases, similarly,  two different sorts of order parameters are necessary, with one residing in the xy-plane and the other locating in the yz-plane (or xz-plane). As has been introduced above, the four orders are order parameters of the instabilities due to the formation of particle-hole bound state and are therefore the EI phases. We term the four EI arising from the $ \hat{\mathbf{e}}_{1} $ sector the EI-1(2a,2b,2c,2d) phases, respectively. Their specific configuration reads respectively as,
\begin{gather}
\boldsymbol{\Delta}^L_{2a}+\boldsymbol{\Delta}^R_{2a}=(\Delta^L_{cz}+\Delta^R_{cz})(\frac{\hat{x}+i\hat{y}}{\sqrt{2} })\notag, \\
\boldsymbol{\Delta}^L_{2b}+\boldsymbol{\Delta}^R_{2b}=(\Delta^L_{pz}+\Delta^R_{pz})\hat{z} \notag, \\
\boldsymbol{\Delta}^L_{2c}+\boldsymbol{\Delta}^R_{2c}=(\Delta^L_{px}+\Delta^R_{px})\hat{x} \notag, \\
\boldsymbol{\Delta}^L_{2d}+\boldsymbol{\Delta}^R_{2d}=(\Delta^L_{cx}+\Delta^R_{cx})(\frac{\hat{y}+i\hat{z}}{\sqrt{2} }).
\end{gather}
Following the standard mean-field procedure, the self-consistent equation of the EI phases are obtained as,
\begin{equation}
1=\frac{g}{2\Omega }\sum _{\mathbf{k}}\frac{\left | \hat{\boldsymbol{\Delta}}\cdot \hat{\mathbf{e}}_1\right |^{2}}{2E_{\mathbf{k}}}\textup{tanh}\frac{\beta E_{\mathbf{k} }}{2}.
\end{equation}
where
\begin{equation}
E_{\mathbf{k}}=\sqrt{\left ( \hbar v_{F} \right )^{2}\left ( k+\omega k_{z} \right )^2+\left |\boldsymbol{\Delta \left ( \mathbf{k} \right )\cdot \hat{e}}_1 \right |^{2}}.
\end{equation}
Last, let us focus on the $ \hat{\mathbf{e}}_{2} $ sector of the intra-node scattering, which is reduced to,
\begin{equation}
\begin{split}
V_3&=-\frac{g}{2\Omega }\sum _{\mathbf{k,k,\pm  }}\left [\hat{\mathbf{e}}_{2} (c^{L\dagger }_{\mathbf{k},\mp }c^{L }_{\mathbf{k},\pm  }-c^{R\dagger }_{\mathbf{k},\mp } c^{R }_{\mathbf{k},\pm  })\right ]^\ast \\
 &\cdot  \left [\hat{\mathbf{e}}_{2}' (c^{L\dagger }_{\mathbf{k'},\mp }c^{L }_{\mathbf{k'},\pm  }-c^{R\dagger }_{\mathbf{k'},\mp } c^{R }_{\mathbf{k'},\pm  })\right ].
\end{split}
\end{equation}
After comparing $V_3$ to $V_2$, the key difference is that the term associated with the excitations around the left and right Weyl points enjoys a minus sign in $V_3$, leading to the following order parameter,
\begin{equation}
\boldsymbol{\Delta}_L+\boldsymbol{\Delta}_R=\frac{g}{2\Omega }\left \langle  \sum _{\mathbf{k'}}\hat{\mathbf{e}}_{2}'
(c^{L\dagger }_{\mathbf{k'},\mp }c^{L }_{\mathbf{k'},\pm  }
-c^{R\dagger }_{\mathbf{k'},\mp }c^{R }_{\mathbf{k'},\pm  }) \right \rangle.
\end{equation}
In the above order parameter, only the vector $\mathbf{\hat{e}_2}$ occurs. Recalling that $\mathbf{\hat{e}_2}$ only spans throughout the the xy-plane and does not include any z-component. Therefore, to describe the symmetry breaking in such order parameter, only two different vectorial configurations are enough to give us all the possible energy minima, with one being the polar phase that locates inside the xy-plane and the other being the chiral phase defined within the xy-plane. We term these two EI orders which arises from the $ \hat{\mathbf{e}}_{2} $ sector as the 3a Polar EI and 3b Chiral EI, respectively.  The corresponding self-consistent equation then can be written as,
\begin{equation}
1=\frac{g}{2\Omega }\sum_{\mathbf{k}}\frac{\left | \hat{ \boldsymbol{\Delta}}\cdot \hat{\mathbf{e}}_2\right |^{2}}{2E_{\mathbf{k}}}\textup{tanh}\frac{\beta E_{\mathbf{k} }}{2}
\end{equation}
where we have set $ \Delta_\alpha =2\Delta^{L/R} _\alpha$ and the energy spectrum is obtained as,
\begin{equation}
E_{\mathbf{k}}=\sqrt{\left ( \hbar v_{F} \right )^{2}\left ( k+\omega k_{z} \right )^2+\left |\boldsymbol{\Delta \left ( \mathbf{k} \right )\cdot \hat{e}}_2 \right |^{2}}.
\end{equation}
%\begin{gather*}
%\boldsymbol{\Delta}^L_{3a}+\boldsymbol{\Delta}^R_{3a}=(\Delta^L_{p}+\Delta^R_{p})\hat{x} \notag \\
%\boldsymbol{\Delta}^L_{3b}+\boldsymbol{\Delta}^R_{3b}=(\Delta^L_{c}+\Delta^R_{c})(\frac{\hat{x}+i\hat{y}}{\sqrt{2}})
%\end{gather*}

Having established all possible eight orders under the effect of interaction. We can solve the eight corresponding and self-consistent equation. All the results are plotted in Fig.5.  Several conclusions can be drawn from Fig.5. First, it is found that the eight orders enjoys similar function dependence between $g$ and $\Delta$. Second, as has been discussed in the CDW case, the trivial solution $(g,\Delta)=(0,0)$ are all present in the eight cases. Third, compared to the type I WSM, all the eight possible mean-field orders are much more easily to be developed, as all the cases have nontrivial solutions down to $g\sim0.1$. Fourth, the polar CDW order (1b) enjoys the larger magnitude of order parameter among all the eight states for any fixed value of $g$. In order to further determine the true ground state among the eight phases, we study the corresponding condensation energy. For $ \mu=0 $, the condensation energy $ E_c $ is defined as the difference between zero temperature free energy of the symmetry breaking orders and that of the normal state.  The results are shown in Fig.6,
\begin{figure}[tbp]
\includegraphics[width=3.9in]{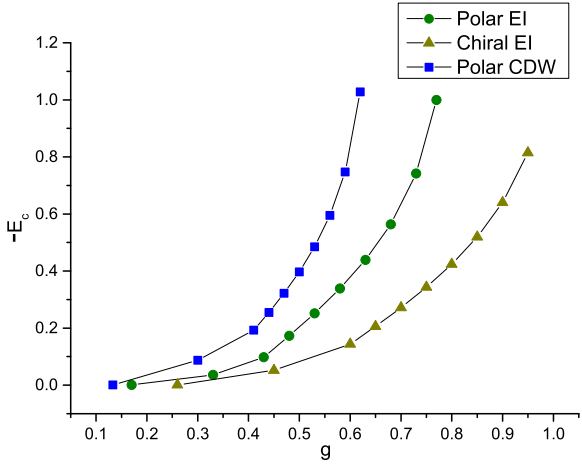}
\caption{The calculated condensation energy (value) of three typical symmetry breaking orders as a function of the interaction $g$.}
\end{figure}
where the magnitude of the condensation energy of the CDW, EI-1, EI-2 states are plotted as a function of the interaction strength (The other states are not energetically favored and are therefore not explicitly shown). From Fig.6, it is found that the polar CDW state enjoys the largest magnitude of  condensation energy among all states. Therefore, we conclude that the polar CDW should be the most likely ground state in interacting type II WSMs. This conclusion is different from that in the type I WSM, where the chiral EI order is found to be more stable. Therefore, this is believed to be a key feature that is unique in the type II case with electron-electron interaction.

\section{a summary}
We have studied the effects of electron-electron interaction in type II WSMs. It is believed that in type I WSM, the interaction can be well neglected since the short-range interaction is irrelevant in RG sense. However, in the type II WSM, we found that the parent state is much more easily to be affected by the short-range interaction, giving rise to several possible instabilities which all breaks the rotation symmetry in the order parameter spontaneously. Through self-consistent mean-field calculation, we established all the possible orders in the particle-hole channel. By comparing their condensation energy, it is found that the polar CDW order arising from the inter-Weyl point scattering is the most likely ground state. The stability of the polar CDW phase is originated from the Fermi surface nesting and therefore enjoys a real space fluctuation period with $\lambda=\pi/k_N$. Since the onset of symmetry breaking orders are much easier than the type I WSM, the observation of the predicted polar CDW phase should be more easier in type II materials with stronger interaction. In addition, our approach can be straightforwardly generalized to other types of topological phases, such as the topological nodal line semimetal. Last, it is also interesting to investigate the role of chiral anomaly after the instabilities set in. If the chiral anomaly remains robust against symmetry breaking, then the resulting phase would enjoy a long-range order in the bulk, together with possible Fermi arc states in the surface. This would be an interesting topic for further study.

%Although our discussion is mainly in the context of the CTI, where a QBCP can serve as the low energy description of its surface state, it is straightforward to make extension to any 2D electron system with nontrivial QBCPs.
%Last, in terms of searching for such hybrid topological insulator, we expect materials that enjoy the checkerboard, Kagome or honeycomb lattice would constitute the promising platforms due to the $C_4$ and $C_6$ rotation symmetry.
\begin{acknowledgments}
We wish to acknowledge Zexun Lin, Jiangtao Yuan,
Bochao Wei, Hangting Chen, Chao Zhang and Jian Wang for fruitful discussions. This work was supported by NSFC Grant No. 11574217 and No. 60825402.
\end{acknowledgments}

\section{Appendix}
\subsection{Calculation of Feynman diagrams}
From Eq.(6) \& Eq.(7), it is not difficult to notice that $ I_A+I_B=0 $, hence we only need to calculate the value of $ I_C $.
\begin{widetext}
\begin{equation}
I_{C}^{(1)}\left ( i\nu ,\mathbf{q} \right )=g^{2}\int \frac{d\omega }{2\pi }\int \frac{d^3p}{ ( 2\pi)^{3}}
\left [  G\left ( \mathbf{k+q-p},i\eta +i\nu-i \omega  \right )G\left ( \mathbf{p+k'},i\omega +i\eta ' \right )\right ]
\end{equation}
\end{widetext}
\begin{widetext}
\begin{equation}
\begin{split}
I_{C}^{(1)}\left ( i\nu ,\mathbf{q} \right )&=g^{2}\int \frac{d\omega }{\pi }\int \frac{d^3p}{ ( 2\pi )^{3}}\frac{-(\eta +\nu- \omega ) (\omega +\eta ' )+v_{F}^{2}\mathbf{(k+q-p)\cdot (p+k')}}{\left [ (\eta +\nu- \omega ) ^{2}+v_{F}^{2}\mathbf{ \left | k+q-p\right |^{2}} \right ]\left [ (\omega +\eta ' )^2 + v_{F}^{2}\mathbf{\left | p+k' \right | }^{2}\right]}\\
&=g^{2}\int \frac{d\omega }{\pi }\int \frac{d^3p}{ ( 2\pi )^{3}}\frac{-(\eta +\nu +\eta ' - \omega ) \omega +v_{F}^{2}\mathbf{(k+q+k'-p)\cdot p}}{\left [ ((\eta +\nu +\eta ' - \omega ) ) ^{2}+v_{F}^{2}\mathbf{ \left |k+q+k'-p\right |^{2}} \right ]\left [ \omega ^2 + v_{F}^{2}\mathbf{\left | p \right | }^{2}\right]}
\end{split}
\end{equation}
\end{widetext}
By using Feynman Tricks :
\begin{equation}
\frac{1}{AB}=\int ^1_0dx\frac{1}{\left [ xA+(1-x)B \right ]^2}
\end{equation}
We have

\begin{equation}
I_{C}^{(1)}\left ( i\nu ,\mathbf{q} \right )=
-g^{2}\int ^1_0dx\int \frac{d\omega }{\pi }\int \frac{d^3p}{ ( 2\pi )^{3}}\frac{- \omega^2 +v_{F}^{2}\mathbf{p}^2+x(1-x)[(\eta +\nu +\eta ')  ^2-v_F^2(\mathbf{ k+k'+q)}^2]}{ \left \{  \omega^2+v^2_F\mathbf{p}^2+x(1-x)\left [ (\nu +\eta +\eta ')^2+v^2_F(\mathbf{q+k+k'})^2 \right ]\right \}^2}
\end{equation}

Here we take a transformation $ \omega \rightarrow \omega -x\nu ,\mathbf{q}\rightarrow \mathbf{q}-x\mathbf{(k+k'+q)} $ and notice the results of special integrals
\begin{widetext}
\begin{equation}
\begin{split}
&\int_{-\infty }^{+\infty }dx\frac{x^2}{(Ax^2+B)^2}=\frac{\pi}{2}\frac{1}{\sqrt{A^3B}}\\
&\int_{-\infty }^{+\infty }dx\frac{1}{(Ax^2+B)^2}=\frac{\pi}{2}\frac{1}{\sqrt{AB^3}}
\end{split}
\end{equation}
\end{widetext}
hence
\begin{widetext}
\begin{equation}
I_{C}^{(1)}\left ( i\nu ,\mathbf{q} \right )=g^2\int_0^1dx\int\frac{d^3p}{(2\pi)^3}\frac{x(1-x)v^2_F(\mathbf{q+k+k'})^2}{\left \{ v^2_F\mathbf{p}^2+x(1-x)\left [ (\nu +\eta +\eta ')^2+v^2_F(\mathbf{q+k+k'})^2 \right ] \right \}^{3/2}}
\end{equation}
\end{widetext}
The integral is restricted to an energy cut-off $ \Lambda $ and a natural momenton cut-off along z-axis $ \tilde \Lambda_0/\hbar v_F $
\begin{equation}
\sqrt{k^2_x+k^2_y+k_z^2}+wk_z=\frac{\Lambda }{\hbar v_F}
\end{equation}
\begin{equation}
\left ( k_z-w\rho  \right )^2-\frac{k_x^2+k_y^2}{w^2-1}=\rho ^2
\end{equation}
set $ \rho =\frac{\Lambda / \hbar v_F }{w^2-1} $and $ -\tilde\Lambda_0 <k_z<\tilde{\Lambda_0} $. Here we have made a hyperbolic transformation.
\begin{gather}
k_z-w\rho =\pm \rho \textup{cosh}t \notag \\
k_x =(w^2-1)^{1/2}\rho \textup{sinh}t\textup{cos}\vartheta \notag \\
k_y =(w^2-1)^{1/2}\rho \textup{sinh}t\textup{sin}\vartheta
\end{gather}
\begin{widetext}
\begin{equation}
I_{C}^{(1)}\left ( i\nu ,\mathbf{q} \right )
=g^2\int_0^1dx\int_{\Lambda' }^{\Lambda }\int_{0}^{t^*}\int_{0}^{2\pi }\frac{x(1-x)(\mathbf{q+k+k'})^2\rho ^2w(w^2-1)\textup{cosh}t\textup{sinh}t}{v_F[(\pm w\textup{cosh}t+1)^2+\Gamma ^2]^{3/2}}d\rho dtd\vartheta
\end{equation}
\end{widetext}
In this equation, $ t^* $ is determined by the boundaries of two branches of hyperbolic curve. After lengthy steps, we finally get
\begin{widetext}
\begin{equation}
\begin{split}
I_{C}^{(1)}\left ( i\nu ,\mathbf{q} \right )&=g^2\int_0^1x(1-x)dx\frac{(\mathbf{q+k+k'})^2}{v_F} \frac{\pi w^2 (w+2) (2 w+3)}{ (w-1) (w+1)^3}\textup{ln}\frac{\Lambda }{\Lambda '} \\
&=g^2\frac{(\mathbf{q+k+k'})^2}{v_F} \frac{\pi w^2 (w+2) (2 w+3)}{ 6(w-1) (w+1)^3}dl \\
&=\lambda g^2 dl
\end{split}
\end{equation}
\end{widetext}
\begin{equation}
g^{\prime}=g+\lambda(\mathbf{k},\mathbf{k}^{\prime},\mathbf{q}) g^2 dl,
\end{equation}
\begin{equation}
\lambda(\mathbf{k},\mathbf{k}^{\prime},\mathbf{q})=\frac{(\mathbf{q+k+k'})^2}{v_F} \frac{\pi w^2 (w+2) (2 w+3)}{ 6(w-1) (w+1)^3}.
\end{equation}
\subsection{Computation of self-consistent equations}
 In Eq.(23), $  t^\ast$ is determined by the boundaries of two branches of hyperbolic curve due to the energy cut-off $ \Lambda $ and a natural momenton cut-off along z-axis $ \tilde \Lambda_0/\hbar v_F $, i.e.
$ -\tilde\Lambda_0/\hbar v_F  <k_z<\tilde{\Lambda_0}/\hbar v_F  $.Explicitly,
if we let
\begin{gather}
\mu =\tilde\Lambda_0/\hbar v_F \rho -w \notag \\
\lambda =\tilde\Lambda_0/\hbar v_F \rho +w
\end{gather}
we have
\begin{enumerate}
	\item if $ \mu>1 $, the integral range $  t^\ast$ have two branches contribution, for one branch, it is \par
	$ \left [\ln\left [ \mu -\sqrt{\mu ^2-1} \right ],
	\ln\left [ \mu +\sqrt{\mu ^2-1} \right ]  \right ] $ \par 
	and \par
	 $ \left [\ln\left [ \lambda -\sqrt{\lambda ^2-1} \right ],
	\ln\left [ \lambda +\sqrt{\lambda ^2-1} \right ]  \right ] $ .
	\item  if $ -1<\mu<1 $, there is only one branch $ \left [\ln\left [ \lambda -\sqrt{\lambda ^2-1} \right ],
	\ln\left [ \lambda +\sqrt{\lambda ^2-1} \right ]  \right ] $ left.
	\item if $ \mu<-1 $, there is also one branch left while it is separated into two parts, i.e.
	$ \left [\ln\left [ -\mu -\sqrt{\mu ^2-1} \right ]\right ],
	\left [\ln\left [ \lambda -\sqrt{\lambda ^2-1} \right ]\right ] $ \par
	and \par
	$ \left [\ln\left [ -\mu +\sqrt{\mu ^2-1} \right ]\right ],
	\left [\ln\left [ \lambda +\sqrt{\lambda ^2-1} \right ]\right ] $ .
\end{enumerate}

\end{document}